# Metamagnetic behavior and effect of field cooling on sharp magnetization jumps in multiferroic $Y_2CoMnO_6$

J. KRISHNA MURTHY [1], K. D. CHANDRASEKHAR[1,2], H. C. WU[2], H. D. YANG[2], J. Y LIN[3] AND A. VENIMADHAV[1 (A)]

[1]*Cryogenic Engineering Centre, Indian Institute of Technology, Kharagpur-721302, India*
[2]*Department of Physics and Center for Nanoscience and Nanotechnology, National Sun Yat-Sen University, Kaohsiung 804, Taiwan*
[3]*Institute of Physics, National Chiao Tung University, Hsinchu 30010, Taiwan*

**Abstract** – We present sharp magnetization jumps and field induced irreversibility in magnetization in multiferroic $Y_2CoMnO_6$. Appearance of magnetic relaxation and field sweep rate dependence of magnetization jumps resemble the martensite like scenario and suggests the coexistence of E*-type antiferromagnetic and ferromagnetic phases at low temperatures. In $Y_2CoMnO_6$, the critical field required for the sharp jump can be increased or decreased depending on the magnitude and direction of the cooling field; this is remarkably different from manganites or other metamagnetic materials where the critical field increases irrespective of the direction of the applied field cooling. The cooling field dependence on the sharp magnetization jumps has been described by considering exchange pinning mechanism at the interface, like in exchange bias model.

**Introduction.** – Metamagnetic materials exhibit a first order irreversible phase transition between two energetically competing magnetic phases. Recently, metamagnetic phase transition has been received a renewed research interest because of its presence in diversified complex magnetic systems. This phenomenon is manifested by sharp jumps in the magnetization with external perturbations like magnetic field [1], temperature [2-4] and pressure [5]. Technological significance is high when the sharp jumps in magnetization are also associated by abrupt changes in other functional properties like, magnetocaloric, magnetostriction and magnetoresistance [6, 7]. Experiments have established that these metamagnetic phase transitions are independent of microstructure and indeed related to the intrinsic nature of the materials [8].

Over the last decade, the magnetic field induced metamagnetic phase transition has been studied extensively in various systems like phase separated manganites [1] and intermetallic alloys such as $Nd_5Ge_3$, $Gd_5Ge_4$ and $CeFe_2$ etc. [9-11]. Recently, metamagnetic behaviour has also been reported in some of the well known multiferroic systems such as $BiFeO_3$ and phase separated multiferroic $Eu_{1-x}Y_xMnO_3$ (x=0.2, 0.25) systems; interestingly, a coupling between metamagnetic behaviour and ferroelectric polarization with the external magnetic field has been noticed [12-14]. Though the exact origin of this effect is unclear, several mechanisms have been proposed in different systems, such as the field dependent orbital ordering in $Pr_{0.5}Ca_{0.5}Mn_{0.95}Co_{0.05}O_3$ [15], martensitic like transformation associated with interface strains in phase separated systems [9-11, 16], spin quantum transition in $Pr_{5/8}Ca_{3/8}MnO_3$ [17], geometric frustration in garnets [18], spin reorientation in FeRh thin films [2] and magnetic field induced spin flop transition in $Ca_3CoMnO_6$ [19]. In charge ordered manganite systems, the field induced magnetization irreversibility with first order nature was assigned to the intrinsic magnetic phase separation, i.e., the coexistence of competing magnetic phases in micro/nano length scales [20], where avalanche-like growth of FM clusters in the vicinity of critical magnetic field ($H_C$) lead to sharp changes in magnetization. On the other hand, in the Heusler alloys the metastability has been ascribed to the interplay of martensitic stains among the crystallographic phases; i.e. a first order structural transition between the low-temperature tetragonal martensite to higher temperature cubic austenite phase [21]. Such systems exhibit characteristics like, (*i*) isothermal field induced sharp magnetization jumps, (*ii*) effect of field cooling on these jumps, and (*iii*) step like growth in magnetic relaxation with respect to time under critical magnetic field and at constant temperature.

A detailed investigation on the effect of field cooling on sharp magnetization jumps can be found in phase separated manganites, and some of the rare earth alloys [10, 15, 16]. Incidentally, in all the investigations, the magnitude of cooling field increases the critical field required for a sharp jump irrespective of the direction of cooling field, and this issue has not been addressed elaborately in the literature.

Recently, improper magnetic multiferroicity was predicated in $Y_2NiMnO_6$ and found experimentally in $Lu_2CoMnO_6$ and $Y_2CoMnO_6$ (YCMO) double perovskite systems, where the E*-type AFM ordering with collinear ↑↑↓↓ spin structure breaks the spatial inversion symmetry and leads to the spontaneous polarization [22, 23, 24]. Such simultaneous

---

existence of both AFM and ferroelectric ordering is of significant interest in data storage and spintronic applications [25, 26]. In this report, we present the field induced sharp magnetization jumps similar to martensite like scenario at low temperatures in YCMO polycrystalline sample. In contrast to other metamagnetic systems, in YCMO, the magnitude of critical field required for a sharp jump can be changed depending on the magnitude and direction of cooling field; this has been described based on the exchange bias and interface pinning mechanism.

**Experimental results and discussion.-** Polycrystalline YCMO sample was prepared by conventional solid state method, crystal structure, and lattice parameters obtained from the Rietveld refinement match well with the previous report (monoclinic crystal structure with space group P $2_1$/n and crystallographic parameters: a = 5.233 Å, b = 5.590 Å, c = 7.470 Å and β = 89.948°) [24]. Temperature and magnetic field dependence of dc susceptibility measurements were done by Quantum Design SQUID-VSM magnetometer. Room temperature and low temperature (~ 10 K) X-ray absorption spectra (XAS) of Co-$L_{2,3}$ and Mn-$L_{2,3}$ data collected at the Dragon beam line of the National Synchrotron Radiation Research Centre in Taiwan with energy resolution of 0.25 eV at the Co-$L_3$ edge (~780 eV).

Fig. 1(a) shows temperature variation of dc magnetization under zero-field-cooled ($M_{ZFC}$) and field-cooled warming ($M_{FCW}$) modes with 0.01 T dc field. A paramagnetic (PM) to FM transition is obtained at 75 K followed by a weak anomaly ~ 55 K related to slow spin dynamics [23].The large irreversibility between FC and ZFC of M (T) data at the onset of magnetic ordering can be related to magnetic anisotropy or glassy behavior. In this regard, we have measured M (T) at different fields (like 1 T, 3 T and 5 T) (not shown here) and magnetic irreversibility  near to ~ 55 K disappears with field; this indicates that observed broad magnetic anomaly is not related to the SG feature [27], it is rather related to the slow dynamics of domain-wall motion [23]. The M *vs*. H curve at 2 K is shown in the Fig. 1(b). It can be seen that the virgin branch of the curve increases with two sharp jumps, one at $H_{C1}$ ~ 1.86 T and other at $H_{C2}$ ~ 2.4 T respectively.  A magnetization of ~ 4.5 $\mu_B$/f.u. at 7 T field has been observed, which is smaller than the theoretically calculated spin only contribution of ~ 6 $\mu_B$/f.u.. However, in Lu$_2$CoMnO$_6$ single crystalline sample [28] the saturation magnetization has been achieved at moderate fields (~ 3 T) in contrasts to its polycrystalline sample, where saturation was found at 60 T. This hints at the important role of magnetic pinning forces on the macroscopic magnetism in polycrystalline samples.

As shown in Fig. 1(b), an irreversibility with respect to first and second branches of M (H) loops (i.e., field induced sharp magnetization jumps are not observed during the second cycle, i.e., H → 0 T case) has been noticed. In fact, irreversibility is one of the common characteristic of metamagnetic systems, yet there is a subtle difference in these behaviours. Based on the nature of the irreversibility behaviour metamagnetic systems can classify into two categories. In type-1, the irreversible loop shows zero remanent magnetization with the strong AFM ground state; Ex: CeFe$_2$ and Pr$_{0.5}$Ca$_{0.5}$Mn$_{1-x}$Co$_x$O$_3$ (x = 0.05) [11, 15]. While in type-2 materials, irreversibility ends up with high remanent magnetization (or permanent transformation to FM state) like Nd$_5$Ge$_3$ [9] and YCMO. Moreover, in YCMO the virgin curve prominently lies outside the M (H) envelope (Fig. 1(b)), and the  metamagnetic phase transition from E*-type AFM to FM state is the first order in nature. The first order metamagnetic phase transition in YCMO indicates the phase coexistence of E*-AFM and FM phases and this complements Sharma *et al.* work, where magnetic inhomogeneity was probed by neutron diffraction [24]. Though metamagnetic behaviour is present up to 10 K, the abrupt nature of field induced transition is apparent only below 4.6 K as shown in Fig. 1(c). Here, the metamagnetic behaviour is a consequence of magnetic phase separation and the phase diagram derived from isothermal magnetization curves is shown in the Fig. 1(d), where the transition from phase separated state to FM state can be realized.

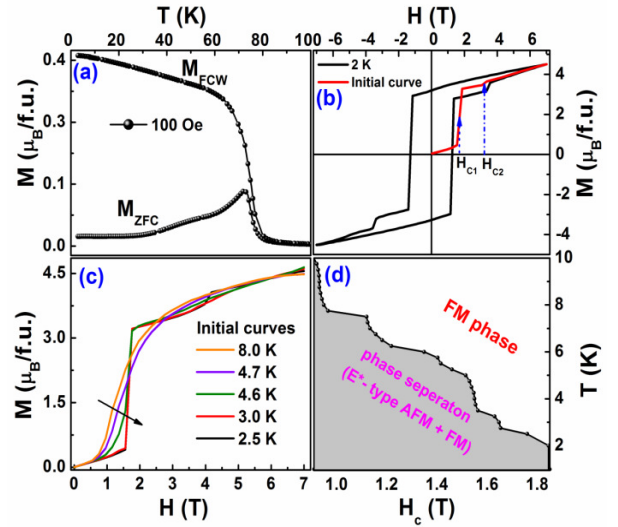

Fig. 1: (Colour on-line) (a) M *vs*. T data in ZFC, FCW mode for 0.01 T field, (b) isothermal M (H) at 2 K, (c) virgin branch of M (H) curves at different temperatures, and (d) temperature dependent magnetic phase diagram obtained from the virgin branch of M (H) curves.

Field induced sharp jumps in multiferroic Ca$_3$CoMnO$_6$ was attributed to the spin flop transition from E*-type magnetic structure (↑↑↓↓) to the ↑↑↑↓ Mn$^{4+}$ (high spin S-3/2) at $H_{C1}$ ~ 11 T and to ↑↑↑↑ with Co$^{2+}$ (low spin S-1/2) spin flop at $H_{C2}$ ~ 25 T fields [19]. Similarly, one can speculate the field induced spin reorientation of Co and Mn ions as the possible origin of the observed metamagnetic behaviour in YCMO. Correspondingly, as shown in the Fig. 1(b), we have observed a change of ~ 2.72 $\mu_B$ in magnetization at the first jump and, which is close to Mn$^{4+}$ (S=3/2) spin flop transition. However, a small change of ~ 0.64 $\mu_B$ noticed at the second jump is not consistent with the spin flop of $Co^{2+}$ (high spin S=3/2) to ↑↑↑↑ magnetic structure. Therefore, field induced spin flop may not be a possible origin for metamagnetic behaviour.

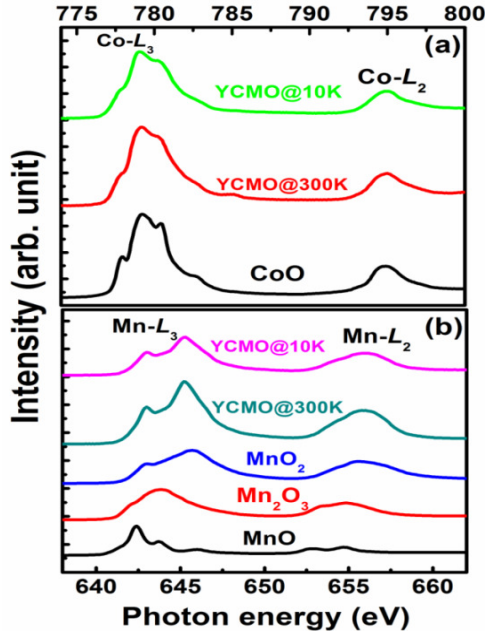

Fig. 2: XAS of YCMO at the (a) Co-$L_{2,3}$ and (b) Mn-$L_{2,3}$ edges at 10 K and 300 K with references of $Co^{+2}O$, $Mn^{+2}O$, $Mn^{+3}_2O_3$ and $Mn^{+4}O_2$ spectra.

Further, we investigate the Co and Mn valence states through the XAS, performed at the Mn and Co-$L$ edges in sample current mode using synchrotron radiation from the 6 m high-energy spherical monochromator (HSGM) beamline at the National Synchrotron Radiation Research Center in Taiwan. In Fig. 2, XAS spectra at 300 K and 10 K are shown to establish the spin states of Co and Mn respectively. As shown in the figure, the Co- $L_{2,3}$ edge peaks (Fig. 2(a)) matches well with the CoO and Mn-$L_{2,3}$ edges (Fig. 2,(b)) lie in the same energy position as $MnO_2$, confirming the divalent state of Co and tetravalent state of Mn respectively. The room temperature spectra of YCMO are similar to that of $LaCo_{0.5}Mn_{0.5}O_3$, and $EuCo_{0.5}Mn_{0.5}O_3$ systems [29, 30]. Further, from Fig. 2 it can be noticed that the spectral line shape and absorption energies are similar for 300 K and 10 K XAS data, which means that temperature has no effect on the valance state of $Co^{2+}$ and $Mn^{4+}$. Moreover, based on neutron diffraction results at 4 K on isostructural $Lu_2CoMnO_6$ reported by Vilar *et al.* it is clearly demonstrated that both $Co^{2+}$ and $Mn^{4+}$ ions are in the high spin state of S= 3/2 [23]. This observation precludes the spin state crossover as the possible source for the observed field induced sharp jumps.

On the other hand, the field induced metamagnetic transition with sharp jumps in M (H) data could be depicted by the martensite like scenario. In YCMO, after cooling the sample in ZFC mode, at low temperature, the E* type AFM ordering would be dominant along with small FM clusters. In fact in most of the metamagnetic manganites such phase separation has been observed with predominant AFM state [31].

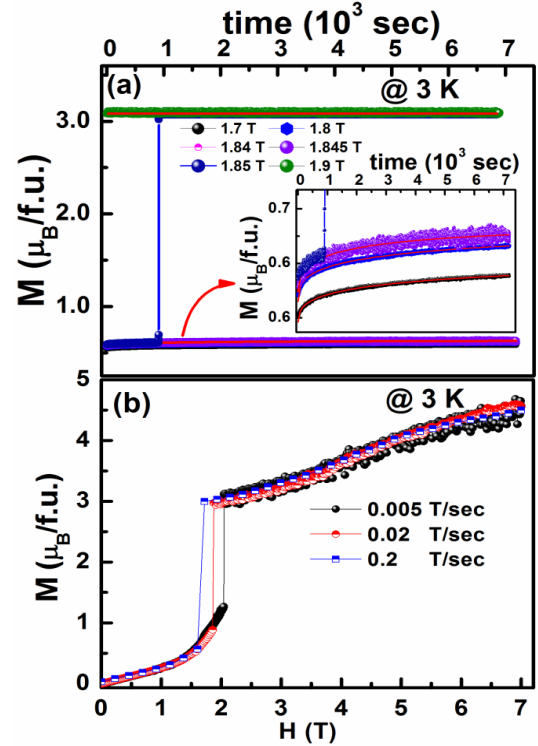

Fig. 3: (a) M vs. time at 3 K under different fields in the vicinity of metamagnetic phase transition (red line represents the stretched exponential function fit to M (t) data ), inset is the magnified view of the M *vs.* time, and (b) isothermal initial M (H) curves at 3 K for different field sweep rates.

The minor phase of FM nucleation sites grow with the increase of the magnetic field during the isothermal magnetization process, consequently there will be continuous change in the interface area between FM and AFM matrix. At a critical field, the Zeeman energy of an external field overcomes the magnetostriction energy related to the strain at the FM/AFM domain interface. Such martensitic strains related structural distortion can induce a burst like growth of FM phase with large magnetization at the expense of AFM domains. And in martensitic like scenario, one can expect magnetic spin relaxation phenomena and ramp rate dependence of sharp jumps as discussed below.

The dynamics of magnetic spin relaxation phenomena in the phase separated YCMO has been done in the vicinity of the critical field. In this protocol, initially the system is brought from PM state to a low temperature (∼ 3 K) under ZFC mode, and then by applying a constant magnetic field, magnetization is allowed to evolve with time. Magnetization as a function of time for different magnetic fields (H) is shown in the Fig. 3(a). Here, for H < 1.845 T, M (t) shows a gradual increase with time (inset to Fig. 3(a)), while for H = 1.85 T, after certain incubation time, a sudden jump in the relaxation curve with high value of magnetization can be noticed. In other words, this is the time required for the applied field energy to overcome the magnetoelastic barrier which appears across the coexisting interfaces [1, 15]. For H >1.85 T, the magnetization is time independent that suggests the complete transformation to FM state. Further, we have fitted this M (t) data (represented with a solid line) to the stretched exponential function of the form, M (H, t) = M (H,0) +[M (H,∞)-M (H, 0)]{1-exp(-t/$\tau$)$^\beta$}, where $\beta$ is the dispersion parameter lies in between 0 and 1 and $\tau$ is the relaxation time for the magnetic spins, which is related to the energy barrier between metastable states. With increasing applied field, the $\tau$ value decreases. A $\tau$ value of ~ 5192 sec at 1.7 T matches with the phase separated manganite systems [32] and it decreases to ~1900 sec in the vicinity of $H_C$ ~ 1.845 T which indicates the decrease of the energy barrier between the metastable states near the phase transition.

We have also studied the magnetization jumps in YCMO by varying sweep rate of the magnetic field. In Fig, 3(b), under ZFC mode, the first cycle of M (H) curve recorded for different applied field sweep rates (0.005, 0.02 and 0.07 T/sec) is shown. For lowest sweep rate, the jump is observed at a critical field ($H_C$) ~ 1.95 T. With the increase of sweep rate, $H_C$ gets shifted to the lower field side and for 0.2 T/Sec, $H_C$ ~ 1.56 T. For lower sweep rates, the lattice has adequate time to adapt the induced strains, while for higher sweep rates, like an impulse, strain propagates rapidly and converts to the FM phase [32]. The magnitude of sweeping field increases the volume fraction of FM phase at the expense of AFM background.

In metamagnetic systems, the critical field required for a sharp jump often increases with field cooling (irrespective of direction of the applied field). In YCMO, we have measured the first and second branches of M (H) loop (with constant sweep rate ~ 0.05 T/s), after cooling the system from PM state to 3 K under different field cooled ($H_{FC}$) conditions. In the Fig. 4(a) the results are shown and compared with the ZFC case (i.e., $H_{FC}$ = 0 T). On cooling the system under different $H_{FC}$, a certain volume fraction of the sample is converted to FM phase correspondingly the initial magnetization ($M_{in}$) value at H = 0 T increases with $H_{FC}$ as shown in the inset of Fig. 4(a). Further, with the increase of $H_{FC}$, $H_c$ shifts towards the higher fields and the variation of both $H_c$ and $M_{in}$ with $H_{FC}$ is nonlinear (inset of Fig 4 (a)). This behaviour is in contrast to manganites where, $H_C$ varies linearly with $H_{FC}$. For $H_{FC} \geq$ 0.14 T the magnetization data does not show a sharp jump; instead a gradual variation with sweeping field is observed. Additionally, we have investigated the dependence of $H_C$ by cooling the sample in the negative fields. Fig. 4(b) shows the first and second branches of M (H) loop for $H_{FC}$ = -0.03, -0.08, -0.1 T. Here, the $H_C$ value is smaller than the ZFC value of 1.845 T and decreases further for higher negative cooling fields. This behaviour is in clear contrast with other metamagnetic systems like $Pr_{0.5}Ca_{0.5}Mn_{0.95}Co_{0.05}O_3$ where $H_C$ was found to increase with the magnitude of $H_{FC}$ and does not depend on the direction of cooling field [15]. It can be understood that for type-1 materials, there is no directional dependence of cooling field, and either direction of applied field would always increase $H_C$. While in type-2 systems, due to the remanent magnetization, there exists a clear FM and AFM interface which is responsible for the directional dependence of $H_C$ with $H_{FC}$. It has been realized that the spin pinning mechanism is likely to be related to the martensitic accommodation of strain across the magnetic interfaces and at $H_C$ the interface overcomes the pinning force by releasing a large strain [33]. Here the cooling field ($H_{FC}$) modulates the interface spin structure, such that it increases or decreases the $H_C$. The effect is interface driven and resembles like exchange bias (EB) phenomena. But the conventional EB effect is absent for obvious reason that for higher fields during M(H) measurement the entire AFM phase changes to FM consequently no interface anisotropy exists. However, one can verify the interface exchange coupling after $H_{FC}$ by performing the minor loop magnetization measurements (with sweeping field < $H_C$ in ZFC condition). As shown in the Fig. 4(c), with the increase of the hysteresis measurement field the FC hysteresis loop shift towards positive field and magnetization values implies the existence of interface exchange coupling. Further, for sweeping fields larger than $\pm$ 1.0 T (i.e., for $\pm$ 5 T as shown in the inset to Fig. 4(c)), the FC-M (H) loop does not show any shift and suggests that interface coupling (or simply FM/FM interface) has disappeared.

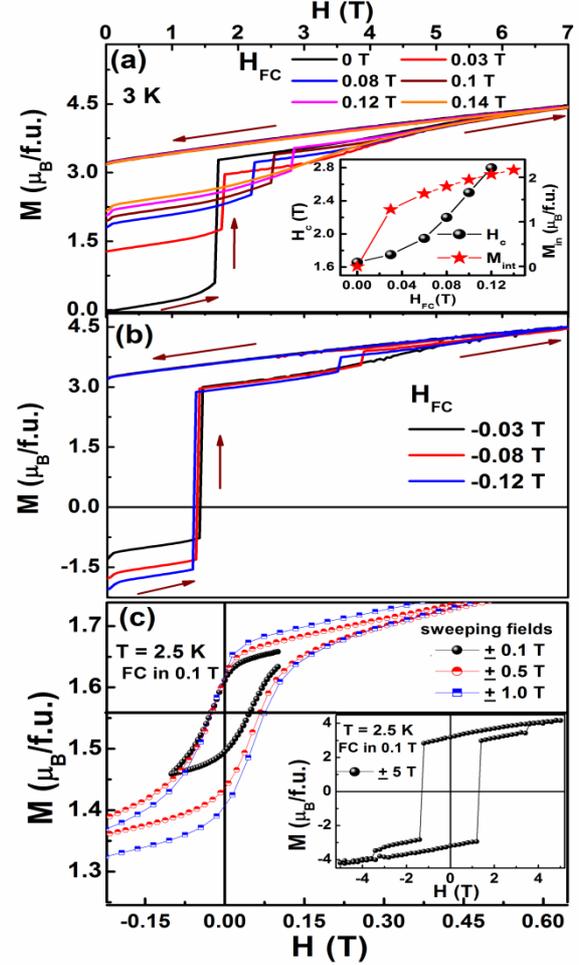

Fig. 4: Initial and second branches of M (H) loop at 3 K under different (a) +$H_{FC}$ and (b) -$H_{FC}$. The arrows indicate the field sweep direction. Inset of (a) shows the variation of critical field ($H_C$), and initial value of magnetization ($M_{in}$) at 3 K for various +$H_{FC}$ and (c) FC-M (H) loops at 2.5 K with different sweeping fields and its inset shows for ± 5 T sweeping fields.

The field cooling dependence of $H_C$ is reminiscent to positive EB effect where it is believed that the interface is AFM coupled (J<0) [34]. We have assumed the exchange coupling J<0 at the FM and AFM interface in YCMO to explain the magnetization jump ($H_C$) in the initial curve with different field cooling conditions ($H_{FC}$). Fig. 5 shows initial M (H) loops after cooling the system in $H_{FC}$ = 0 and $\pm$ 0.1 T and schematics of one FM/AFM interface with their initial interface states at H = 0 T and final state (at H = 7 T). Here, 'A' depicts the spin state for $H_{FC}$ = 0 with dominant E*-AFM state with randomly oriented FM clusters. During the first branch, FM clusters start to grow in volume adjacent to the E*-AFM neighbour with large pinning force. At $H_C$, the Zeeman energy overcomes the pinning force, and magnetization jumps due to burst like growth of FM clusters at the expense of AFM domains and converts to FM ordering as depicted in 'F' state. The scenario is different for the field cooled case, where there exist remanent magnetization due to the partial conversion to FM phase due to increasing number of FM nucleation sites and their size with $H_{FC}$ and hence a clear FM/E*-AFM interface with negative exchange coupling (J <0) such that the interface spin structure is aligned antiferromagnetically with the FM neighbourhood. Now, the $H_C$ depends on the sign and strengths of the FM layer, interface spin structure and the external magnetic field. For $H_{FC}$ = +0.1 T, as shown in 'B' the FM layer is already in the direction of the applied magnetic field while the interface is negatively coupled, and this leads to the interfacial exchange or pinning energy to the total magnetoelastic energy that restrains the magnetization jump. To overcome such a total oppositional force one need to apply more Zeeman energy in terms of external field that means the shifting of the metamagnetic sharp jump towards the high field side. With the increase of $H_{FC}$, the FM phase grows in volume, and the interface pinning energy also increases, and such a situation leads to a higher critical field $H_C$. For $H_{FC} \geq$ 0.14 T, the induced FM phase dominates as evidenced from the large $M_{in}$ (=2.1$\mu_B$/f.u.) and magnetization shows a smooth variation. While for $H_{FC}$ = -0.1 T, the FM spins are in the opposite direction of the sweeping magnetic field (evident from the negative value of the $M_{in}$), while the pinned interface spins in the AFM region is in the direction of the applied field as shown in 'C' and this favours the magnetization jump. The $H_C$ value depends mainly on the FM- Zeeman energy, but this may not vary with the $H_{FC}$. However, with increasing the magnitude of -$H_{FC}$, the critical field shifts towards lower fields due to the proportional increase in the area of the interface that favours the jump. Hence, the induced exchange pinning across the interface of FM/AFM phases is responsible for the shifting of the sharp jump across the metamagnetic phase transition. The model can also be valid for the type-I case, in particular to mangnaites where there is an evidence of glassy phase [35]. And this glassy phase recovers after the re-moval of magnetic field and the system is back to situation like 'A' state either before or after the field cooling. However, the number of nucleation sites increases with the magnitude of the field cooling and hence it increases $H_C$ but remain independent of direction of the cooling field.

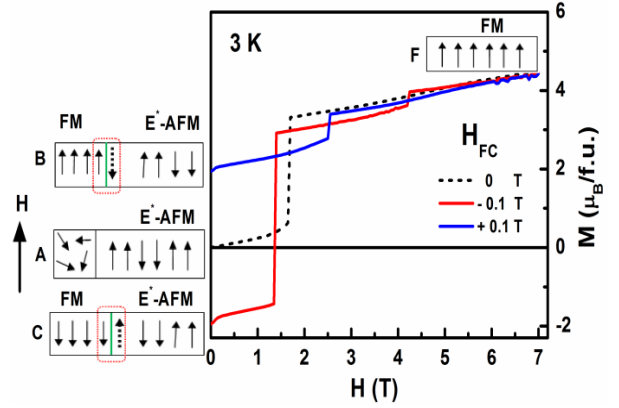

Fig. 5: Schematic representation of the resultant interface spin configuration to illustrate the initial and final magnetic states after cooling the system under 0 T and $\pm$ 0.1 T.

**Conclusions.** - In conclusion, the field induced sharp magnetization jumps at low temperatures and magnetization irreversibility with large remanence suggests the magnetic phase separation in YCMO. The time-dependent magnetic relaxation, field sweep rate and field cooled dependence of sharp jumps are consistent with the martensitic scenario and suggest such a field induced phase transition from the E*-type AFM and FM ordering is a first order in nature. We find that the critical field can be increased or decreased depending on the direction of field cooling. The dependence of $H_C$ on the magnitude and direction of field cooling reveals the role of interface exchange pinning like in exchange bias model.

\*\*\*

The authors acknowledge to IIT Kharagpur for funding VSM-SQUID magnetometer. Krishna, thanks CSIR-UGC, Delhi for SRF. This work was partially supported by National Science Council, Taiwan under Grant No. NSC 100-2112-M-110-004-MY3. We are indebted to NSRRC, Hsinchu, Taiwan for doing the XAS measurements.

J. Krishna Murthy *et al.*

REFERENCES

[1] HARDY V., MAIGNAN A., HÉBERT S., YAICLEe C., MARTIN C., HERVIEU M., LEES M. R., ROWLANDS G., PAUL D. Mc. K. and RAVEAU B., *Phys. Rev. B,* **68** (2003) 220402.
[2] BORDEL C., JURASZEKURASZEK J., COOKE D. W., BALDASSERONI C., MANKOVSKY S., MINAR J., EBERT H., MOYERMAN S., FULLERTON E. E. and HELLMAN F., *Phys. Rev. Lett.*, **109** (2012) 117201.
[3] SWOBODA T. J., CLOUD W. H., BITHER T. A., SADLER M. S. and JARRETT H. S., *Phys. Rev. Lett.,* **4** (1960) 509.
[4] GRATZ E., MARKOSYAN A. S., GAIDUKOVA I. YU., RODIMIN V. E., BERGER ST., BAUER E. and MICHOR H., *Solid State Commun.,* **120** (2001) 191.
[5] ARSLANOV T. R., MOLLAEV A. YU., KAMILOV I. K., ARSLANOV R. K., KILANSKI L., TRUKHAN V. M., CHATTERJI T., MARENKIN S. F. and FEDORCHENKO I. V., *Appl. Phys. Lett.,* **103** (2013) 192403.
[6] TRIGUERO C., PORTA M. and PLANES A., *Phys. Rev. B,* **76** (2007) 094415.
[7] MATSUKAWA M., YAMATO Y., KUMAGAI T., TAMURA A., SURYANARAYANAN R., NIMORI S., APOSTU M., REVCOLEVSCHI A., KOYAMA K. and KOBAYASHI N., *Phys. Rev. Lett.*, **98** (2007) 267204.
[8] OUYANG Z. W., NOJIRI H. and YOSHII S., *Phys. Rev. B,* **78** (2008) 104404.
[9] MAJI B., SURESH K. G. and NIGAM A. K., *Euro phys. Lett.,* **91** (2010) 37007.
[10] VELEZ S., HERNANDEZ J. M., FERNANDEZ A., MACIÀ F., MAGEN C., ALGARABEL P. A., TEJADA J. and CHUDNOVSKY E. M., *Phys. Rev. B,* **81** (2010) 064437.
[11] ROY S. B., CHATTOPADHYAY M. K., CHADDAH P. and NIGAM A. K., *Phys. Rev. B,* **71** (2005) 174413.
[12] TOKUNAGA M., AZUMA M. and SHIMAKAWA Y., *J. Phys. Soc. Jpn.*, **79** (2010) 064713.
[13] DANJOH S., JUNG J. –S., NAKAMURA H., WAKABAYASHI Y. and KIMURA T., *Phys. Rev. B,* **80** (2009) 180408(R).
[14] CHOI Y. J., ZHANG C. L., LEE N. and CHEONG S. –W., *Phys. Rev. Lett.,* **105** (2010) 097201.
[15] MAHENDIRAN R., MAIGNAN A., HÉBERT S., MARTIN C., HERVIEU M., RAVEAU B., MITCHELL J. F. and SCHIFFER P., *Phys. Rev. Lett.*, **89** (2002) 286602.
[16] HARDY V., MAJUMDAR S., CROWE S. J., LEES M. R., PAUL D. MC. K., HERVÉ L., MAIGNAN A., HÉBERT S., MARTIN C., YAICLE C., HERVIEU M. and RAVEAU B., *Phys. Rev. B,* **69** (2004) 020407.
[17] CAO G., ZHANG J., CAO S., JING C. and SHEN X., *Phys. Rev. B,* **71** (2005) 174414.
[18] TSUI Y. K., BURNS C. A., SNYDER J. and SCHIFFER P., *Phys. Rev. Lett.*, **82** (1999 ) 3532.
[19] FLINT R., YI H. T., CHANDRA P., CHEONG S. –W. and KIRYUKHIN V., *Phys. Rev. B,* **81** (2010) 092402.
[20] DAGOTTO E., 2002 N*anoscale phase separation and colossal magnetoresistance* (berlin: springer) (www.springer.de), UEHARA M., MORI .S, CHEN C. H. AND CHEONG S. –W., *Nature,* **399** (1999) 560.
[21] ROY S. B., *J. Phys: Condens. Matter,* **25** (2013) 183201.
[22] SANJEEV K., GIANLUCA G., JEROEN V. D. B. and SILVIA P., *Phys. Rev. B,* **82** (2011) 134429.
[23] YÁÑEZ-VILAR S., MUN E. D., ZAPF V. S., UELAND B. G., GAR DNER J. S., THOMPSON J. D., SINGLETON J., SÁNCHEZ-ANDÚJAR M., MIRA J., BISKUP N., SEÑARÍS –R. M. and BATISTA C. D., *Phys. Rev. B,* **84** (2011) 134427.
[24] SHARMA G., SAHA J., KAUSHIK S. D., SIRUGURI V. and PATNAIK S., *Appl. Phys. Lett.,* **103** (2013) 012903.
[25] SPALDIN N. A. and FIEBIG M., *Science,* **309** (2005) 391.
[26] CHEONG S. –W. and MOSTOVOY M., *Nat. Mater*, **6** (2007) 13.
[27] KRISHNA MURTHY J. and VENIMADHAV A., *J. Phys. D: Appl. Phys.,* **Accepted, (in press)**.
[28] LEE N., CHOI H. Y., JO Y. J., SEO M. S., PARK S. Y. and CHOI Y., J *Appl. Phys. Lett.,* **104** (2014) 112907.
[29] BURNUS T., HU Z., HSIEH H. H., JOLY V. L. J., JOY P. A., HAVERKORT M. W., WU H., TANAKA A., LIN H. –J., CHEN C. T. and TJENG L. H., *Phys. Rev. B,* **77** (2008) 125124.
[30] VASILIEV A. N., VOLKOVA O. S., LOBANOVSKII L. S., TROYANCHUK I. O., HU Z., TJENG L. H., KHOMSKII D. I., LIN H. J., CHEN C. T., TRISTAN N., KRETZSCHMAR F., KLINGELER R. and BÜCHNER B., *Phys. Rev. B,* **77**(2008) 104442.
[31] DEAC I. G., MITCHELL J. F., and SCHIFFER P., *Phys. Rev. B,* **63**(2001) 172408.
[32] LIAO D. Q., SUN Y., YANG R. F., LI Q. A. and CHENG Z H., *Phys. Rev. B,***74** (2006) 174434.
[33] NIEBIESKIKWIAT D. and SÁNCHEZ R. D., *J. Phys: Condens. Matter,* **24** (2012) 436001.
[34] NOWAK U., USADEL K. D., KELLER J., MILTÉNYI P., BESCHOTEN B. and GÜNTHERODT G., *Phys. Rev. B,* **66** (2002) 014430.
[35] ZHOU S. Y., LANGNER M. C., ZHU Y., CHUANG T. –D., RINI M., GLOVER T. E., HERTLEIN M. P., GONZALEZ C. A. G., TAHIR N., TOMIOKA Y., TO- KURA Y., HUSSAIN Z. and SCHOENLEIN R. W., *Sci. Reports*, **4** (2014) 4050.